\documentclass{aa}
\pdfoutput=1
\usepackage{amsmath}
\usepackage[round]{natbib}
\usepackage{graphicx}
\usepackage{flushend}
\usepackage{paralist}
\usepackage[english]{babel}
\usepackage{txfonts}
\usepackage[pdftitle={Deconvolution with Shapelets},pdfauthor={Peter Melchior et al.},bookmarks=true,bookmarksnumbered=true]{hyperref}
\bibpunct{(}{)}{;}{a}{}{,} 

\newcommand\veqref[1]{Eq. \eqref{#1}}
\newcommand\veqsref[2]{Eqs. \eqref{#1} \& \eqref{#2}}
\newcommand\Veqsref[3]{Eqs. \eqref{#1}, \eqref{#2} \& \eqref{#3}}

\title{Deconvolution with Shapelets}
\author{P. Melchior\inst{1} \and R. Andrae\inst{1} \and M. Maturi\inst{1} \and M. Bartelmann\inst{1}}
\institute{Zentrum f\"ur Astronomie, ITA, Universit\"at Heidelberg,
  Albert-Ueberle-Str. 2, 69120 Heidelberg, Germany\\
\email{pmelchior@ita.uni-heidelberg.de}}
\date{}
\abstract
{
}
{
  We seek to find a shapelet-based scheme for deconvolving galaxy images from the PSF which leads to unbiased shear measurements.
}
{
  Based on the analytic formulation of convolution in shapelet space, we construct a procedure to recover the unconvolved shapelet coefficients under the assumption that the PSF is perfectly known. Using specific simulations, we test this approach and compare it to other published approaches.
}
{
  We show that convolution in shapelet space leads to a shapelet model of order $n_{max}^h = n_{max}^g + n_{max}^f$ with $n_{max}^f$ and $n_{max}^g$ being the maximum orders of the intrinsic galaxy and the PSF models, respectively. Deconvolution is hence a transformation which maps a certain number of convolved coefficients onto a generally smaller number of deconvolved coefficients. By inferring the latter number from data, we construct the maximum-likelihood solution for this transformation and obtain unbiased shear estimates with a remarkable amount of noise reduction compared to established approaches. This finding is particularly valid for complicated PSF models and low $S/N$ images, which renders our approach suitable for typical weak-lensing conditions.
}{}

\begin{document}
\thispagestyle{empty}
\maketitle

\section{Introduction}

Shapelets have been proposed as an orthonormal set of two-dimensional functions to quantify shapes of galaxy images \citep{shapeletsI}. They have several convenient mathematical properties suggesting their use in measurements of weak gravitational lensing \citep{shapeletsII}. Their application to data with low signal-to-noise ratio, however, is hampered by a number of problems. First, their enormous flexibility allows shapelets to represent random noise patterns well, which generates their tendency to escape into high orders and to overfit noisy images. Second, this same property not only affects the order, but also the scale of the best-fitting shapelet model. The scales of the shapelets decomposing noisy images is thus likely too high.
These drawbacks are inherent properties of the shapelet method and have to be addressed for an effective usage in astronomical image processing. At first sight, it seems natural to limit the order of the shapelets from above to avoid overfitting, and to limit also the scale from above to prevent shapelets from creeping into the noise. 
On the other hand, convolution with the PSF alters the shape and spatial extent of imaged objects, and thus leaves an imprint on the maximum shapelet order and the scale size.

Guided by these considerations, we address the following question: How should the maximum order and the spatial scale of a shapelet decomposition be determined in cases where PSF convolution significantly modifies the object's appearance? This question aims at applications of shapelets in weak-lensing measurements, where the situation is particularly delicate because typically very small and noisy images are convolved with structured, incompletely known PSF kernels with scales similar to those of the images. How can deconvolution schemes be constructed in this case in order to find significant and unbiased shear estimates?

We show in Sect. \ref{sec:convolution} and in the Appendix that mathematical sum rules exist for the shapelet orders and the squared spatial scales of original image, kernel, and convolved image. We then proceed in Sect. \ref{sec:deconvolution} to devise an algorithm respecting these sum rules as well as possible, which leads to a deconvolution scheme based on the (possibly weighed) pseudo-inverse of a rectangular rather than the inverse of a quadratic convolution matrix. In Sect. \ref{sec:benchmark}, we demonstrate by means of simulations with different signal-to-noise levels and PSF kernels that our algorithm does indeed perform very well, and leads in most realistic cases to substantially improved results compared to previously proposed methods. Our conclusions are summarized in Sect. \ref{sec:conclusions}.

\section{Convolution in shapelet space}
\label{sec:convolution}
A two-dimensional function $f(\mathbf{x})$, e.g. a galaxy image, is decomposed into a set of shapelet modes by projection,
\begin{equation}
\label{eq:decomposition}
f_\mathbf{n} = \int_{-\infty}^\infty d^2x\ f(\mathbf{x})\ B_\mathbf{n}(\mathbf{x};\alpha),
\end{equation}
where $\mathbf{n} = (n_1,n_2)$ is a two-dimensional index and $\alpha$ is called \emph{scale size}. The
two-dimensional shapelet basis function
\begin{equation}
\label{eq:shapelet_basis}
B_\mathbf{n}(\mathbf{x};\alpha) = \alpha^{-1} \phi_{n_1}(\alpha^{-1} x_1)
\ \phi_{n_2}(\alpha^{-1}x_2),
\end{equation}
is related to the one-dimensional Gauss-Hermite polynomial
\begin{equation}
\label{eq:dimless_basis}
\phi_{n}(x) = [2^n \pi^{\frac{1}{2}} n!]^{-\frac{1}{2}}\ H_n(x)\ 
\mathrm{e}^{-\frac{x^2}{2}},
\end{equation}
with $H_n(x)$ being the Hermite polynomial of order $n$.
From the coefficients one can then reconstruct a shapelet model
\begin{equation}
\label{eq:model}
\tilde{f}(\mathbf{x}) = \sum_\mathbf{n}^{n_{max}} f_\mathbf{n}\, B_\mathbf{n}(\mathbf{x};\alpha).
\end{equation}
The number of shapelet modes -- often expressed in terms of the maximum shapelet order $n_{max} = max(n_1,n_2)$ -- and the scale size have to be determined by an optimization algorithm \citep{shapeletsIII,Melchior07.1}, which minimizes the modulus of the residuals $f-\tilde{f}$, or from empirical relations based on other measures of the object like FWHM or major and minor axes \citep{Chang04.1, Kuijken06.1}.

A convolution 
\begin{equation}
\label{eq:convolution}
h(\mathbf{x}) \equiv (f\star g)(\mathbf{x}) \equiv
\int_{-\infty}^\infty d^2 x'
f(\mathbf{x}')g(\mathbf{x}-\mathbf{x}').
\end{equation}
can be performed analytically in shapelet space.
According to \veqref{eq:decomposition}, the functions $f$, $g$ and $h$ are represented by sets of shapelet states $f_\mathbf{n}$, $g_\mathbf{n}$ and $h_\mathbf{n}$ with scale sizes $\alpha$, $\beta$ and $\gamma$.

\cite{shapeletsI} and \cite{shapeletsII} showed that the coefficients of the convolved image $h(\mathbf{x})$ are given by
\begin{equation}
\label{eq:shapelet_convolution}
h_\mathbf{n} = \sum_{\mathbf{m},\mathbf{l}}
C_{\mathbf{n},\mathbf{m},\mathbf{l}}(\alpha,\beta,\gamma) f_\mathbf{m} g_\mathbf{l} = \sum_\mathbf{m} P_{\mathbf{n},\mathbf{m}}(\alpha,\beta,\gamma) f_\mathbf{m}
\end{equation}
where $P_{\mathbf{n},\mathbf{m}} \equiv \sum_\mathbf{l} C_{\mathbf{n},\mathbf{m},\mathbf{l}} g_\mathbf{l}$ is called \emph{convolution matrix}. The value of $C_{\mathbf{n},\mathbf{m},\mathbf{l}}(\alpha,\beta,\gamma)$ can be computed analytically.

\begin{figure}[t]
\includegraphics[width=\linewidth]{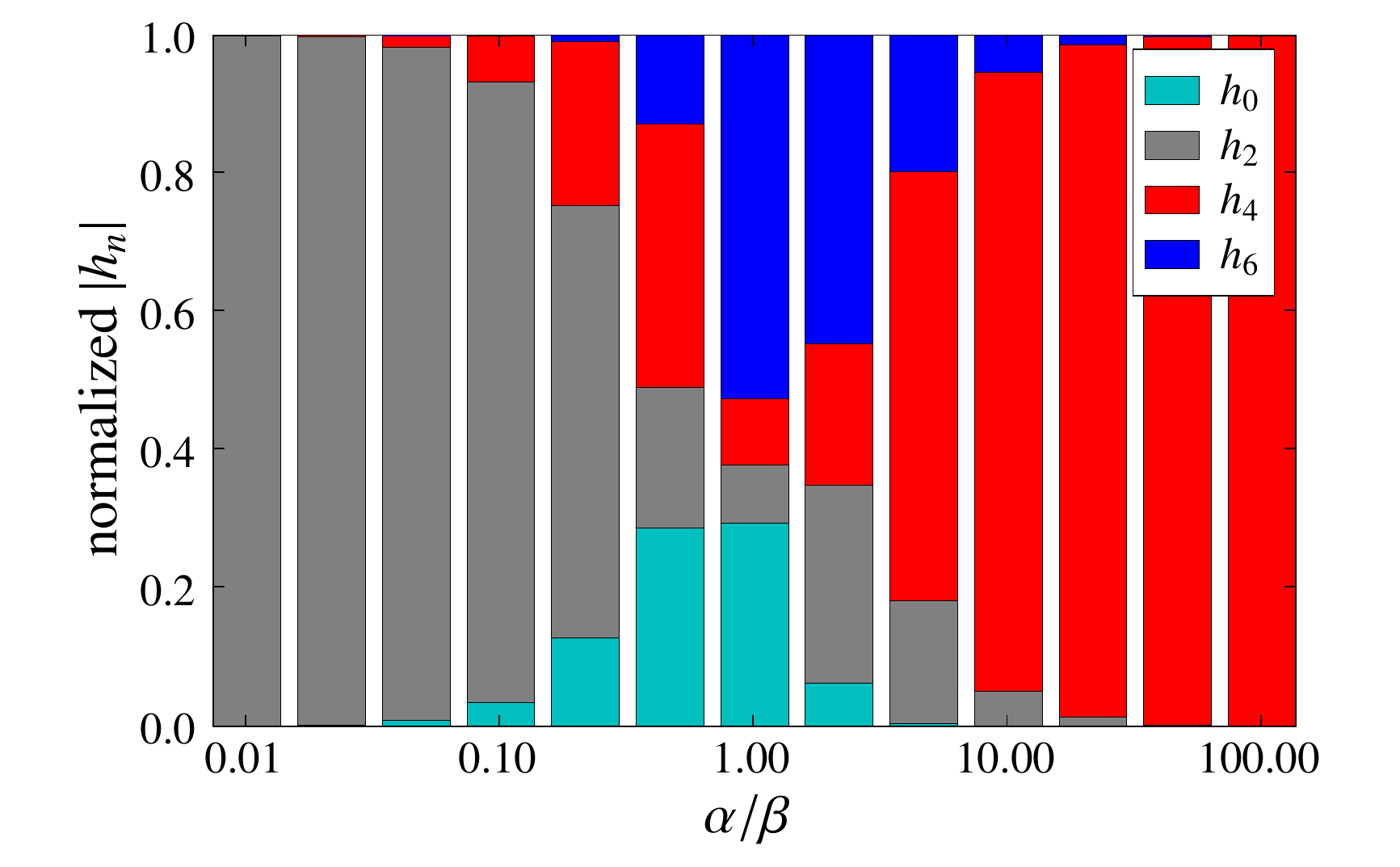}
\caption{One-dimensional convolution $h = B_4(x;\alpha) \star B_2(x;\beta)$. We plot the modulus of $h_n$ with $n$ even (all odd modes have vanishing power), normalized by $\sum_n |h_n|$.} 
\label{fig:conv4_1}
\end{figure}

However, there is no clear statement on the scale size $\gamma$ and, in particular, on the maximum order $n_{max}^h$ of the convolved object $h$. In appendix \ref{sec:proof} we proof that the so-called \emph{natural choice} \citep{shapeletsI}
\begin{equation}
\label{eq:natural_choice}
\gamma^2 = \alpha^2 + \beta^2
\end{equation}
is indeed the correct choice for $\gamma$ and that the maximum order of the convolved object is given by
\begin{equation}
\label{eq:conv_order}
n_{max}^h = n_{max}^f + n_{max}^g.
\end{equation}
While this result gives the highest possible mode of the convolved object which could contain power, it does not tell us whether it does indeed have power, as this depends primarily on the ratio of scales $\alpha/\beta$ entering $P_{\mathbf{n},\mathbf{m}}$. This is demonstrated in Fig. \ref{fig:conv4_1}, where we show the result of a convolution of a function which is given by a pure $B_4$ mode with a kernel represented by a pure $B_2$ mode. From this it becomes obvious that in a wide region around $\alpha/\beta\simeq 1$ power is transfered to all even modes up to $n=6$  (odd modes vanish because of parity, cf. \veqref{eq:conv_final} and the following discussion). If either $\alpha\gg\beta$ or $\beta\gg\alpha$, the highest order of the larger object is also the highest effective order of the convolved object. Thus, we can generalize \veqref{eq:conv_order},
\begin{equation}
\label{eq:effective_conv_order}
 n_{max}^h = 
\begin{cases} 
n_{max}^f\ (+1) & \alpha \gg \beta\, \ \text{(kernel negligible)}\\
n_{max}^f + n_{max}^g &  \alpha\simeq\beta\\
n_{max}^g\ (+1) & \alpha \ll \beta\, \ \text{(kernel dominant)}
\end{cases}
\end{equation}
where the option $(+1)$ is taken if required by parity.

For most cases, in particular for weak gravitational lensing, PSF and object scales are comparable, which means that we must not neglect the power transfer to higher modes.

\section{How to deconvolve}
\label{sec:deconvolution}
While the correct values for $\gamma$ and $n_{max}^h$ are clarified now, it is not obvious how these results need to be used in deconvolving real data. We first comment on possible ways to undo the convolution and then discuss our finding in the light of measurement noise.

\subsection{Deconvolution strategies}
As \citet{shapeletsII} have already discussed, there are two ways to deconvolve from the PSF in shapelet space:
\begin{compactitem}
\item Inversion of the convolution matrix: According to \veqref{eq:shapelet_convolution}, one can solve for the unconvolved coefficients, 
\begin{equation}
\label{eq:P_inversion}
f_\mathbf{m} = \sum_\mathbf{n} P_{\mathbf{m},\mathbf{n}}^{-1}\, h_\mathbf{n}.
\end{equation}
\item Fitting with the convolved basis system \citep{Kuijken99.1,shapeletsIII}, which modifies \veqref{eq:model} such that it directly minimizes the residuals of $h(\mathbf{x})$ w.r.t. its shapelet model
\begin{equation}
\label{eq:PSF_model}
\tilde{h}(\mathbf{x}) = \sum_\mathbf{n}^{n_{max}} f_\mathbf{n}\sum_\mathbf{m}\, P_{\mathbf{n},\mathbf{m}} B_\mathbf{m}(\mathbf{x};\alpha).
\end{equation}
\end{compactitem}

The second method is generally applicable but slow because the convolution has to be applied at each iteration step of the decomposition process. On the other hand, the first approach reduces deconvolution to  a single step after the shapelet decomposition and is therefore computationally more efficient.

According to \veqref{eq:conv_order}, $P$ is not quadratic and thus not invertible as suggested by \veqref{eq:P_inversion}. In order to cope with this, we need to replace the inverse $P^{-1}$ by the pseudo-inverse $P^\dagger\equiv \bigl(P^T P\bigr)^{-1} P^T$ such that the equation now reads
\begin{equation}
\label{eq:P_pseudo_inv}
f_\mathbf{m} = \sum_\mathbf{n} P_{\mathbf{m},\mathbf{n}}^{\dagger}\, h_\mathbf{n}.
\end{equation}

What seems as a drawback at first glance is effectively beneficial. Conceptually, this is now the least-squares solution of \veqref{eq:shapelet_convolution}, recovering the most probable unconvolved coefficients from the set of noisy convolved coefficients. The underlying assumption of Gaussian noise in the coefficients $h_\mathbf{n}$ holds for the most usual case of background-dominated images for which the pixel noise is Gaussian. 

Correlations among the coefficients $h_\mathbf{n}$, which arise from  non-constant pixel weights or pixel correlations, can be accounted for by introducing the coefficient covariance matrix $\Sigma\equiv\langle(h_\mathbf{n} - \langle h_\mathbf{n} \rangle)(h_\mathbf{n} - \langle h_\mathbf{n} \rangle)^T\rangle$, which alters \veqref{eq:shapelet_convolution} to read
\begin{equation}
P^T \Sigma P f = P^T \Sigma h.
\end{equation}
Maximizing the likelihood for recovering the correct unconvolved coefficients leads to the weighted pseudo-inverse
\begin{equation}
P^\dagger_w\equiv (P^T \Sigma P)^{-1} P^T \Sigma,
\end{equation}
which replaces $P^\dagger$ in \veqref{eq:P_pseudo_inv}.

However, both approaches (direct inversion, \veqref{eq:P_inversion}, or least-squares solution, \veqref{eq:P_pseudo_inv}) would fail if $P$ was rank-deficient. \citet{shapeletsII} argued that convolution with the PSF amounts to a projection of high-order modes onto low-order modes and therefore $P$ can become singular. This is true only for very simple kernels (e.g. the Gaussian-shaped mode of order 0) with rather large scales. In fact, \veqref{eq:effective_conv_order} tells us that convolution carries power from all available modes of $f$ to modes up to order $n_{max}^h\geq n_{max}^f$, hence $P$ is generally not rank-deficient. In practice we did not have problems in constructing $P^{-1}$ or $P^\dagger$ when using realistic kernels. We therefore see no hindrance in employing the matrix-inversion scheme and will use it in the course of this paper.

\subsection{Measurement process and noise}
Up to here, we have discussed (de-)convolution entirely in shapelet space, where this problem is now completely solved. For the following line of reasoning, we will further assume that the kernel is perfectly known and can be described by a shapelet model. 

Critical issues still arise at the transition from pixel to shapelet space: There are no intrinsic values of $n_{max}^f$ and $\alpha$, and even if they existed they would not directly be accessible to a measurement. While the first statement stems from the fact that we try to model a highly complicated galaxy or stellar shape with a potentially completely inappropriate function set, the second statement arises from pixelation and measurement noise occurring in the detector.

However, the pixelated version of the shape can be described by a shapelet model, with an accuracy which depends on the noise level and the pixel size. Consider for example a galaxy whose light distribution strictly follows a S\'ersic profile. Modeling the cusp and the wide tails of this profile with the shapelet basis functions would require an infinite number of modes. But pixelation effectively removes the central singularity of the S\'ersic profile and turns the continuous light distribution into a finite number of light measures, such that it is in principle describable by a finite number of shapelet coefficients. Pixel noise additionally limits the spatial region within which the tails of S\'ersic profile remain noticeable and hence the number of required shapelet modes.

Consequently, shapelet implementations usually determine $n_{max}$ by some significance measure of the model \citep[$\chi^2$ in][]{shapeletsIII, Melchior07.1} or -- similarly -- fix $n_{max}$ at a value which seems reasonable to capture the general features of the shape \citep[e.g.][]{shapeletsII, Kuijken06.1}. 

\begin{figure}[t]
\includegraphics[width=\linewidth]{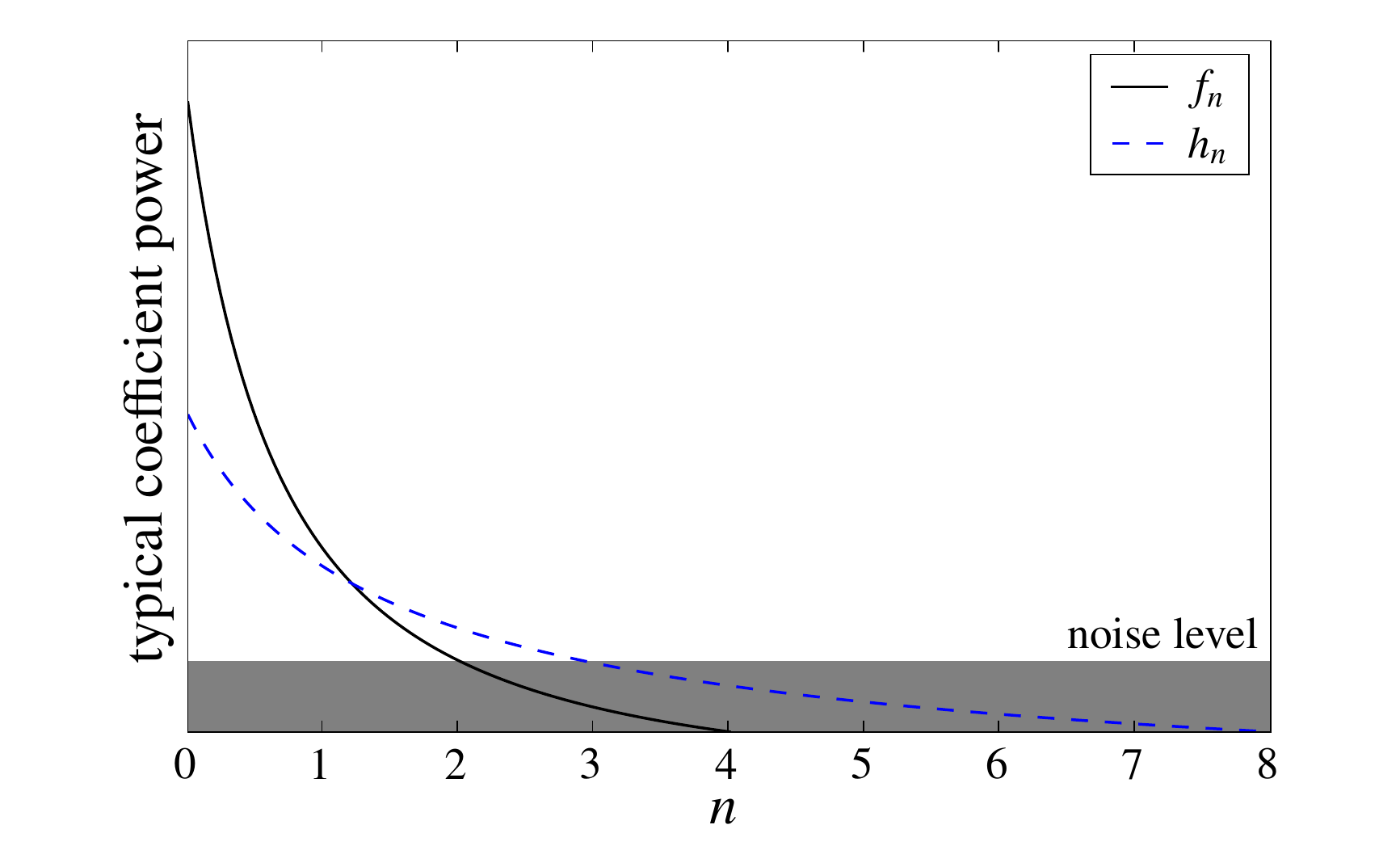}
\caption{Sketch of the effect of a convolution on the power of shapelet coefficients. The detailed shape of the curves is neither realistic nor important, but typical shapelet models show decreasing coefficient power with increasing order $n$. As convolution does not change the overall power of an object but distributes it over more coefficients, the average $S/N$ of shapelet coefficients is lowered. The noise regime (represented by the gray area) is constant in the case of uncorrelated noise.}
\label{fig:sketch}
\end{figure}

Fig. \ref{fig:sketch} schematically highlights an important issue of a significance-based ansatz:
When the power in a shapelet coefficient is lower than the power of the noise, it is considered insignificant, and the shapelet series is truncated at this mode (in Fig. \ref{fig:sketch}, $f_n$ may be limited to $n\leq2$ and $h_n$ to $n\leq3$).
Since convolution with a flux-normalized kernel does not change the overall flux or  -- as the shapelet decomposition is linear -- the total coefficient power but generally increases the number of modes, the signal-to-noise ratio $S/N$ of each individual coefficient is lowered on average. Thus, after convolution more coefficients will be considered insignificant and therefore disregarded. 

This is equivalent to the action of a convolution in pixel space, where some objects' flux is distributed over a larger area. If the noise is independent of the convolution, demanding a certain $S/N$ threshold results in a smaller number of significant pixels.

The main point here is that we try to measure $h_n$ from data and from this $f_n$ by employing \veqref{eq:P_pseudo_inv}. But if we truncate $h_n$ too early -- at an order $n_{max}^h \ll n_{max}^f + n_{max}^g$ --, the resulting unconvolved coefficients $f_n$ are expected to be biased even if the convolution kernel is perfectly described. The reason for this is that by truncating we assume that any higher-order coefficient is zero on average while in reality it is non-zero, it is just smaller than the noise limit. Every estimator formed from these coefficients is thus likely biased itself.

In turn, if we knew $n_{max}^f$, we could go to the order demanded by \veqref{eq:conv_order} and the deconvolution would map many noise-dominated high-order coefficients back onto lower-order coefficients. This way, we would not cut off coefficient power and our coefficient set would remain unbiased. Unfortunately, this approach comes at a price: Firstly, the resulting shapelet models are often massively overfitted, and secondly, obtaining unbiased $f_n$ requires the knowledge of $n_{max}^f$. The first problem can be addressed by averaging over sufficiently many galaxies, while the second one can indeed be achieved by checking the $S/N$ of the recovered $f_n$ \emph{after} deconvolution. The average number of significant deconvolved coefficients gives an indication of the typical complexity of the imaged objects as they would be seen in a measurement without convolution but using the particular detector characterized by its pixel size and noise level.

\subsection{Unbiased deconvolution method}
\label{sec:method}

The previous consideration guides us to set up a deconvolution procedure which yields unbiased deconvolved coefficients. Again, we assume perfect knowledge of the kernel $g$ in shapelet space.
\begin{compactitem}
\item Given the noise level and the pixel size of the images, we initially guess $\bar{n}_{max}^f$
\item We set the lower bounds $n_{max}^h \geq n_{max}^g + \bar{n}_{max}^f$ and $\gamma\geq\beta$.
\item We decompose each galaxy by minimizing the decomposition $\chi^2$ under these constraints. A value of $n_{max}^h > n_{max}^g + \bar{n}_{max}^f$ is used only if $\chi^2>1$ otherwise. This yields $h_\mathbf{n}$ and $\gamma$.
\item By inverting \veqref{eq:natural_choice}, we obtain $\tilde\alpha$.
\item Using the maximum orders and scale sizes for $f$, $g$ and $h$ in addition to $g_\mathbf{n}$, we can form the convolution matrix $P$ according to \veqref{eq:shapelet_convolution}.
\item By forming $P^\dagger_{(w)}$ and applying \veqref{eq:P_pseudo_inv}, we reconstruct $\tilde{f}_\mathbf{n}$.
\item By propagating the coefficient errors from the decomposition through the same set of steps, we investigate the number of significant coefficients and should find $\bar{n}_{max}^f$ if our initial guess was correct.
\end{compactitem}

Given the demanded accuracy, it might be necessary to adjust the guess $\bar{n}_{max}^f$ and reiterate the steps above. For this approach, it is inevitable to split the data set in magnitude bins as the best value for $\bar{n}_{max}^f$ clearly depends on the intrinsic brightness. Further splitting (according to apparent size or brightness profile etc.) may be advantageous, too.

\section{Deconvolution microbenchmark}
\label{sec:benchmark}
There exists a growing number of shapelet-based decomposition and deconvolution approaches published in the literature. In this section we will show that the method proposed here is indeed capable of inferring unbiased unconvolved coefficients. Moreover, employing the least-squares solution given by \veqref{eq:P_pseudo_inv} results in a considerable noise reduction, which is to be expected from this ansatz.

At first, we want to emphasize that the simulations we use in this section are highly simplistic. Their only purpose is to investigate how well a certain decomposition/deconvolution scheme can recover the unconvolved coefficients. By understanding the performance of different approaches, we acquire the knowledge for treating more realistic cases.

\subsection{The testbed}

\begin{figure}[t]
 \includegraphics[width=\linewidth]{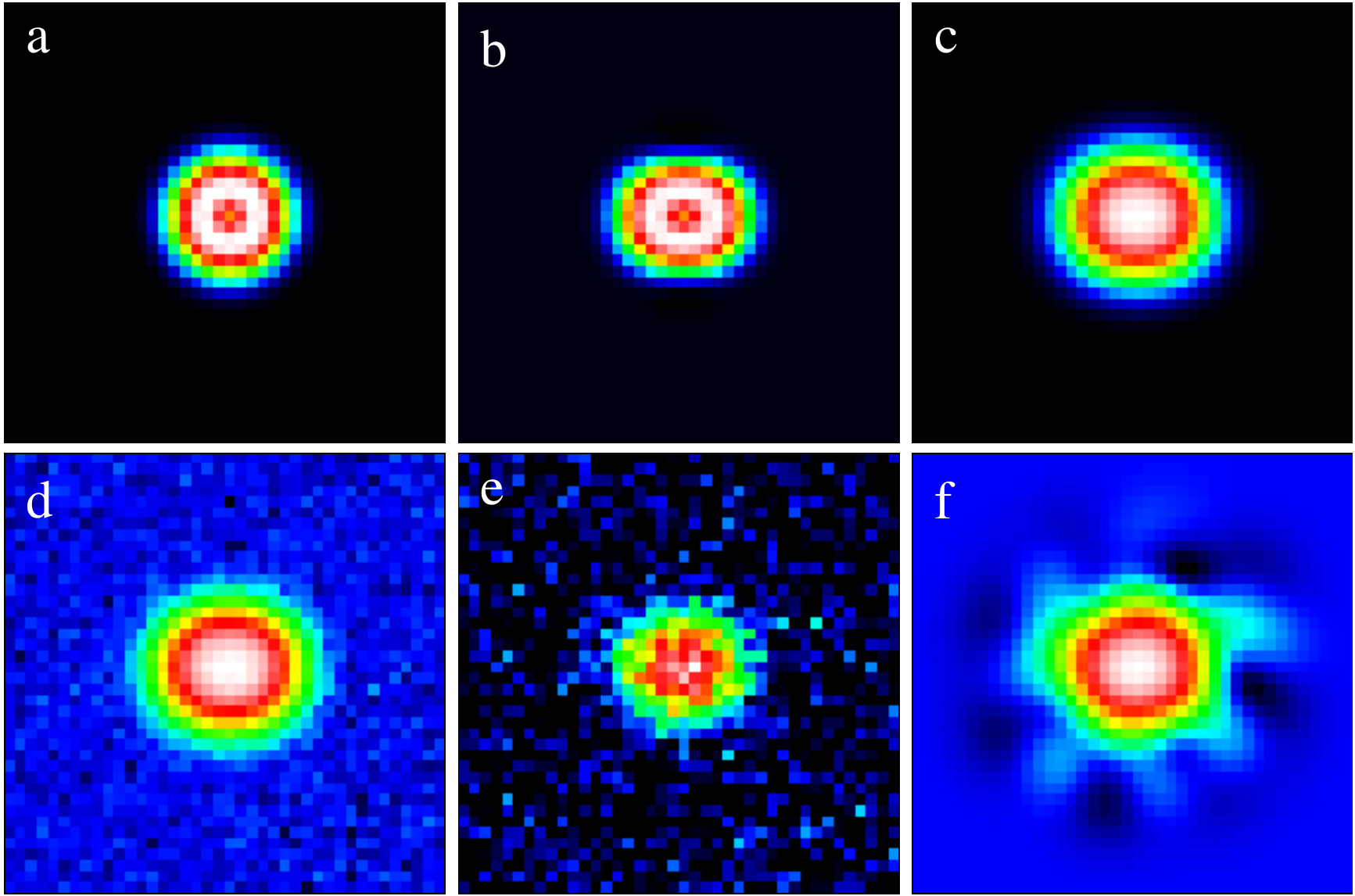}
 \caption{Example of simulated galaxies used in our testbed: (a) intrinsic galaxy model with $\alpha=2$ and flux equaling unity, (b) after applying a shear $\vec{\gamma} = (0.1,0)$, (c) after convolving with PSFb from Fig. \ref{fig:models_kernels} with $\beta=2$. The bottom panels show (c) after addition of Gaussian noise of zero mean and variance $\sigma_n^2$: (d) moderate noise, $\sigma_n = 10^{-4}$, (e)  high noise, $\sigma_n = 10^{-3}$. (f) is the shapelet reconstruction of (e). Colors have logarithmic scaling.}
 \label{fig:models_noise}
\end{figure}

\begin{figure}[t]
 \includegraphics[width=\linewidth]{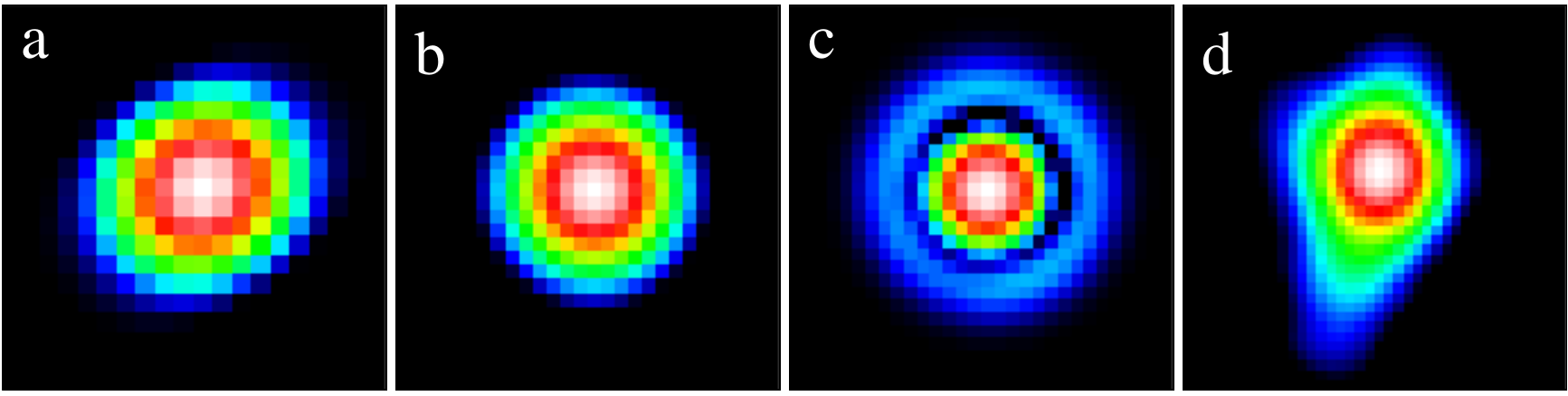}
 \caption{The kernels used in our benchmark: (a) model of PSF2 from STEP1 \citep{STEP1} with $n_{max}^g=4$, (b) model of PSF3 from STEP1 with $n_{max}^g=4$; (c) Airy disk model with $n_{max}^g=6$; (d) model from a raytracing simulation of a space-bourne telescope's PSF with $n_{max}^g=8$ and $n_{max}^g=12$ (shown here). Colors have logarithmic scaling.}
 \label{fig:models_kernels}
\end{figure}

The construction of simulated galaxy images is visualized in Fig. \ref{fig:models_noise}. As intrinsic function we use a polar shapelet model with $f_{0,0}=f_{2,0}=c$, where $c$ is chosen such that the model has unit flux. $\alpha$ is varied between 1.5 and 4. Given its ring-shaped appearance, this model is not overly realistic but also not too simple, and circularly symmetric. We apply a mild shear of $\vec{\gamma} = (0.1,0)$, thus populate coefficients of order $\leq4$, and convolve with five different realistic kernels $g$ (cf. Fig. \ref{fig:models_kernels}) in shapelet space (employing \veqsref{eq:natural_choice}{eq:conv_order} with $1.5\leq\beta\leq6$). The pixelated version of the convolved object is then subject to $N$ realizations of Gaussian noise with constant variance.

Each of these simulated galaxy images is decomposed into shapelets again, yielding $h_\mathbf{n}$, using the code by \citet{Melchior07.1}, where the optimization is constrained by fixing either $n_{max}^h$ or $\gamma$, or both. $h_\mathbf{n}$ is then deconvolved from the kernel $g$. 

As a diagnostic for the correctness of the deconvolved coefficients, we estimate the gravitational shear from the quadrupole moments $Q_{ij}$ of the light distribution \citep{weak-lensing-review},
\begin{equation}
 \label{eq:shear_quad}
 \tilde{\vec{\gamma}}_Q = \frac{1}{2}\ \vec{\chi} \equiv \frac{1}{2}\frac{Q_{11}-Q_{22}+2iQ_{12}}
   {Q_{11}+Q_{22}}\text{,}
\end{equation}
where $Q_{ij}$ is computed as a linear combination of all available deconvolved coefficients \citep{shapelets_manual,Melchior07.1}.

We investigate five different approaches which differ in the choice of $n_{max}^h$, $\tilde{n}_{max}^f$ or the reconstruction of $\alpha$. The different choices are summarized in Tab. \ref{tab:methods}.

\begin{table}[t]
\caption{Overview of the parameter choices of the investigated methods. $n_{max}^h$ is the order of the decomposed object, $\tilde{\alpha}$ the estimate on the intrinsic scale and $\tilde{n}_{max}^f$ an estimate on the intrinsic order of $f$.}
\label{tab:methods}
\centering
\begin{tabular}{lccc}
Name & $n_{max}^h$ & $\tilde{\alpha}$ & $\tilde{n}_{max}^f$ \\
\hline\hline
{\sc Full} & $\geq n_{max}^g + \bar{n}_{max}^f$ & $\sqrt{\gamma^2 - \beta^2}$ & $\bar{n}_{max}^f$\\
{\sc Signific} & $\geq n_{max}^g$ & $\sqrt{\gamma^2 - \beta^2}$ & $\bar{n}_{max}^f$\\
{\sc Same} & $n_{max}^g$ & $\sqrt{\gamma^2 - \beta^2}$ & $n_{max}^g$\\
{\sc ConstScale} & $n_{max}^g$ & $\gamma$ & $n_{max}^g$\\
{\sc Nmax2} & $2$ & $\sqrt{\gamma^2 - \beta^2}$ & $2$\\
\hline
\end{tabular}
\end{table}

{\sc Full} is the method we propose here (cf. Sect. \ref{sec:method}); for the following tests, we set $\bar{n}_{max}^f = n_{max}^f = 4$. 
{\sc Signific} is a variant of {\sc Full}, which bounds the decomposition order by the kernel order because coefficients beyond that are often insignificant, but makes use of our guess on $\bar{n}_{max}^f$.

{\sc Same} is similar to the one used by \citet{Kuijken06.1} with two differences: As discussed above, we employ the matrix inversion scheme (\veqref{eq:P_inversion} since $P$ is square for this method) instead of fitting the convolved shapelet basis functions, and in our implementation $\chi^2$ is minimized w.r.t. a continuous parameter $\gamma$, while \citet{Kuijken06.1} finds the best-fitting $\gamma = 2^{n/8}\beta$ with some integer $n$. Without knowing the increase of shapelet orders due to convolution given by \veqref{eq:conv_order}, this represents the best-defined deconvolution approach.

 \cite{shapeletsII} stated that the approach {\sc ConstScale} delivers the best results in their analysis. {\sc Nmax2}, however, is an approach inspired by the na\"ive assumption that such a decomposition scheme catches the essential shear information without being affected by overfitting.

\subsection{Performance with moderate noise}
The first set of simulations comprise galaxy models with peak $S/N$ between 45 and 220 with a median of $\approx90$ (an example is shown in Fig. \ref{fig:models_noise}d); for each value of $\alpha$ and $\beta$ we created $N=100$ noise realizations. These high $S/N$ values are more typical for galaxy morphology studies rather than for weak lensing, but we can see the effect of the convolution best. In this regime, problems with the deconvolution method become immediately apparent.

\begin{figure}[t]
 \includegraphics[width=\linewidth]{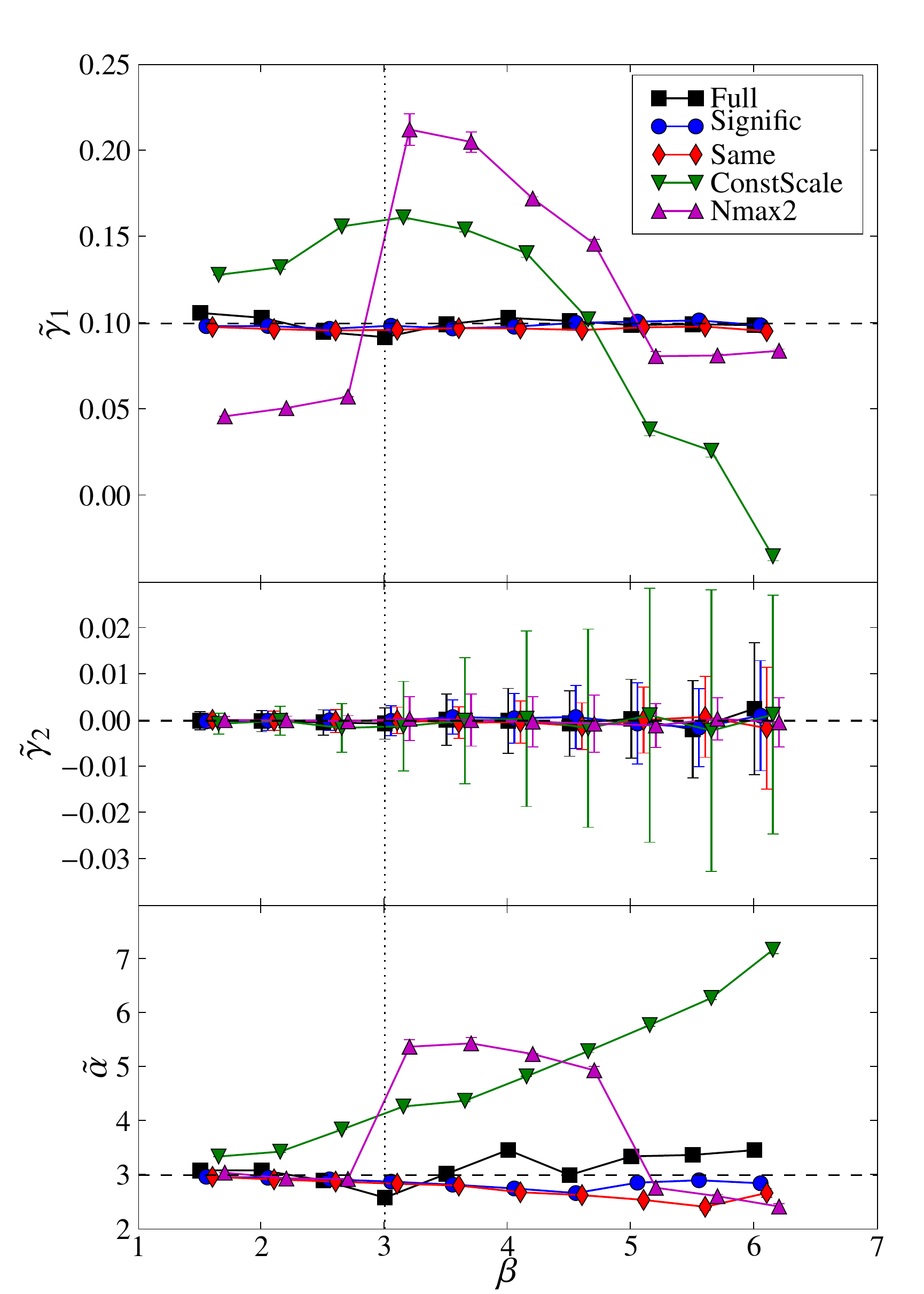}
 \caption{Recovered shear $\tilde{\vec\gamma}$ and intrinsic scale size $\tilde\alpha$ in a moderate-noise simulation ($\sigma_n=10^{-4}$, $\alpha=3$, PSFb) as functions of the kernel scale $\beta$. In each panel the horizontal dashed line shows the true value of the quantity and the vertical dotted line shows the true value for $\alpha$ as reference. Errorbars (which are often too small to be visible) exhibit the standard deviation of the mean of $N=100$ realizations. For visualization purposes, each method is slightly offset along $\beta$. Simulations with different PSF models or $\alpha$ are qualitatively equivalent.}
 \label{fig:methods_low-noise}
\end{figure}

Considering Fig. \ref{fig:methods_low-noise}, we can ascertain that {\sc Full}, {\sc Signific} and {\sc Same} perform quite well while {\sc ConstScale} and {\sc Nmax2} are clearly in trouble. This is not too surprising: By construction, {\sc Nmax2} truncates the shapelet series at $n_{max}^h=2$ and hence misses all information contained in higher-order coefficients. One has to recall that the sheared model already has $n_{max}^f = 4$, after convolution with PSFb ($n_{max}^g=4$) it arrives at $n_{max}^h=8$. {\sc Nmax2} tries to undo the deconvolution with less information than contained in both sheared model and kernel individually. This is an enormously underconstrained attempt and leads to unpredictable behavior. 
{\sc ConstScale} assumes that $\alpha$ can be approximated by $\gamma$ and hence $\tilde{\alpha}$ is almost a increasing function of $\beta$ (see bottom panel of Fig. \ref{fig:methods_low-noise}). According to \veqref{eq:natural_choice}, this ansatz is only applicable if $\beta$ is negligible. For very small kernel scales, we can indeed see a tendency to converge to the correct solution, but for all other situations, this choice is manifestly non-optimal. Because of the clearly problematic behavior of {\sc ConstScale} and {\sc Nmax2}, we exclude these two methods from the further investigation.

This situation is very similar for other choices of $\alpha$ and other PSF models. To work out the general trends of the three remaining methods, we average over all scales $\alpha$ and $\beta$ and plot the results in dependence of the PSF model. 

\begin{figure}[t]
 \includegraphics[width=\linewidth]{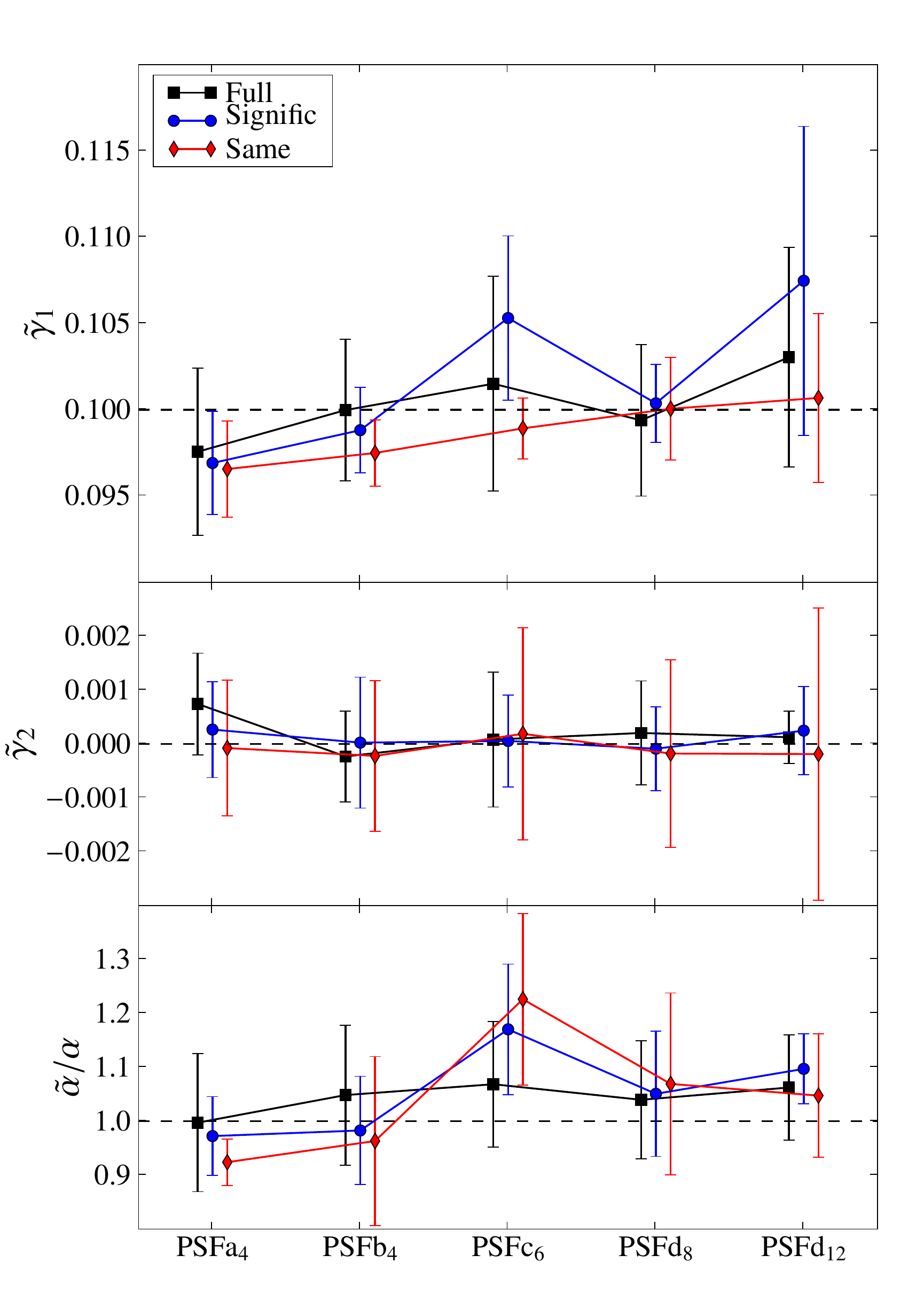}
 \caption{Recovered shear $\tilde{\vec\gamma}$ and intrinsic scale size $\tilde\alpha$ (in units of the true scale size $\alpha$) in the moderate-noise simulations in dependence of the PSF models from Fig. \ref{fig:models_kernels} (subscripts denote $n_{max}^g$). Each data point represents the mean of the quantity for all available values of $\alpha$ and $\beta$ (in total 60 independent combinations), errorbars show the standard deviation of the mean. For visualization purposes, each method is slightly offset horizontally w.r.t. the others.}
 \label{fig:moderate-noise_average}
\end{figure}

The top and middle panels of  Fig. \ref{fig:moderate-noise_average} confirm that all remaining methods yield essentially unbiased estimates of the shear, although we notice a mild tendency of {\sc Same} and {\sc Signific} to underestimate $\vec{\gamma}_1$. This indicates that truncation of the decomposition order $n_{max}^h = n_{max}^g$ might be insufficient for high $S/N$ images. The fact that this underestimation is absent at higher kernel orders confirms this interpretation.

Within the errors, the recovered scale size $\tilde{\alpha}$ is rather unbiased (see the bottom panel of \ref{fig:moderate-noise_average}). For {\sc Same} and {\sc Signific}, we can see a clear shift of $\tilde{\alpha}$ for PSFc. The reason for this lies in the large spatially extent and wide wings of the Airy disk model in combination with a low $n_{max}^h$. Since the entries of $P$ depend in a non-linear way on $\alpha$, this shift affects the recovery of the shear and leads to slightly poorer results.

From this initial simulation with moderate noise we can conclude that one should respect \veqref{eq:natural_choice} and must not truncate the shapelet series of $h_\mathbf{n}$ severely.

\subsection{Performance with high noise}

\begin{figure}[t]
 \includegraphics[width=\linewidth]{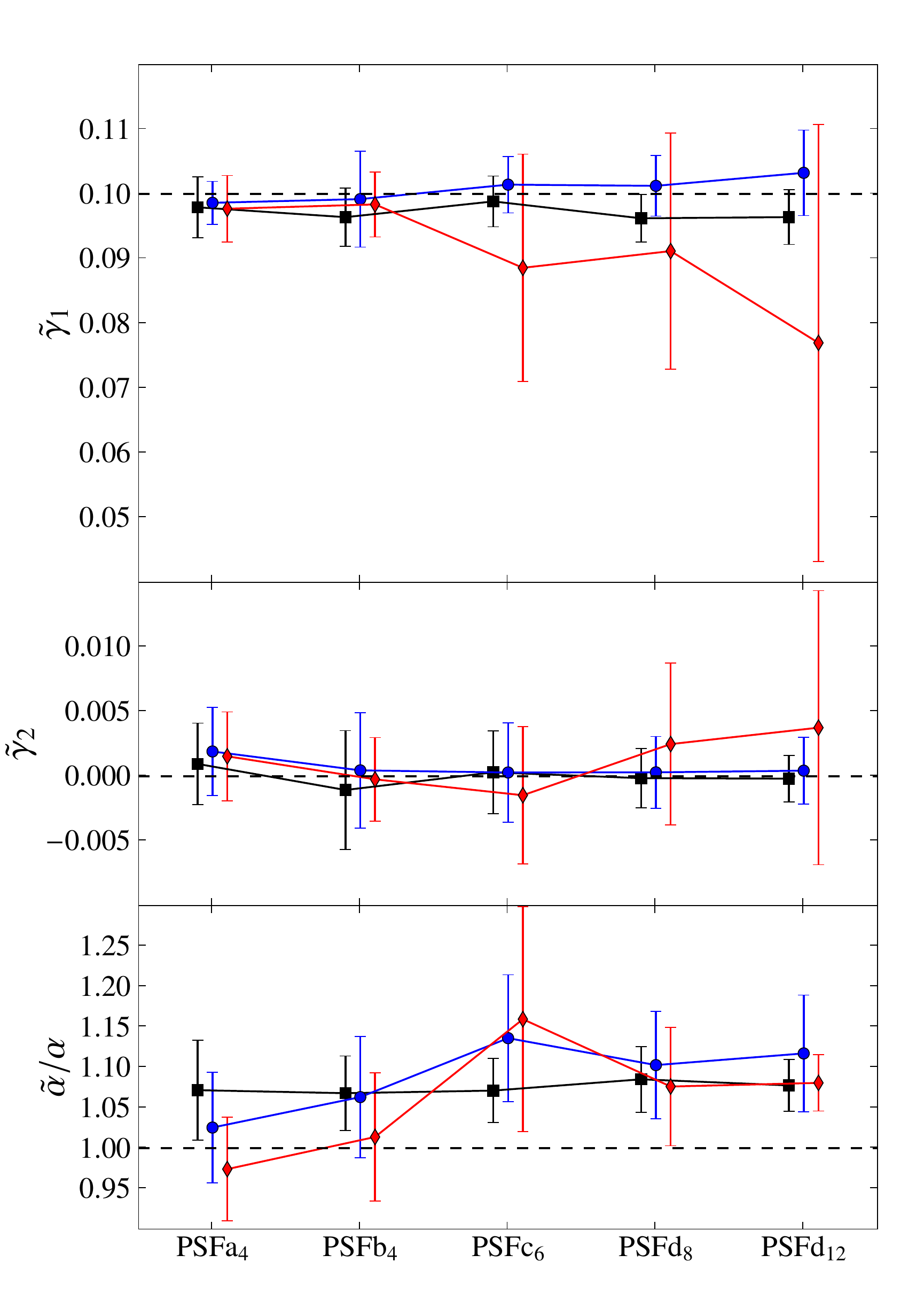}
 \caption{Analogous to Fig. \ref{fig:moderate-noise_average} but for the high-noise simulations.}
 \label{fig:high-noise_average}
\end{figure}

We now consider a realistic weak-lensing situation by increasing the noise level by a factor of 10, hence $4.5\leq S/N\leq 22$ (cf. Fig. \ref{fig:models_noise}e). To balance the increased noise, we also increase the number of realizations to $N=1000$.

Considering Fig. \ref{fig:high-noise_average}, we can confirm that also for very noisy images the shear estimates from these three methods are not significantly biased. However, for {\sc Same} we can see a remarkable drop of the mean of $\vec{\gamma}_1$ and a drastic increase of the noise in $\vec{\gamma}_1$ and $\vec{\gamma}_2$ with the kernel order. Both findings are probably related to the usage of $P^{-1}$ instead of $P^\dagger$ when performing the deconvolution.
In contrast to the two methods we are proposing here, {\sc Same} uses $\tilde{n}_{max}^f = n_{max}^g$ (cf. Table \ref{tab:methods}). For the typical weak-lensing scenario -- characterized by $n_{max}^f < n_{max}^g$, where all methods create a substantial amount of overfitting, cf. Fig. \ref{fig:models_noise}f --, this assumes to find a higher number of significant deconvolved coefficients then are actually available. These additional, noise-dominated coefficients impact on $Q_{ij}$ and $\vec{\gamma}_Q$ (cf. \veqref{eq:shear_quad}), therefore these quantities become rather noisy themselves. Given the fact that those high-order coefficients contain mostly arbitrary pixel noise which does not have a preferred direction, they also tend to dilute the available shear information from the lower-order coefficients, which explains the drop in $\vec{\gamma}_1$. The estimate for $\vec{\gamma}_2$ is not affected as its true value was zero anyway.

The superior behavior of {\sc Full} and {\sc Signific} in these low $S/N$ simulations can also be seen more directly. As measure of the decomposition quality, we calculate the distance in shapelet space between the mean deconvolved coefficients $\tilde{f}_\mathbf{n}$ and the true input coefficients $f_\mathbf{n}$,
\begin{equation}
R_s^2 = \sum_\mathbf{n}\bigl(\langle\tilde{f}_\mathbf{n}\rangle - f_\mathbf{n}\bigr)^2.
\end{equation}

\begin{figure}[t]
 \includegraphics[width=\linewidth]{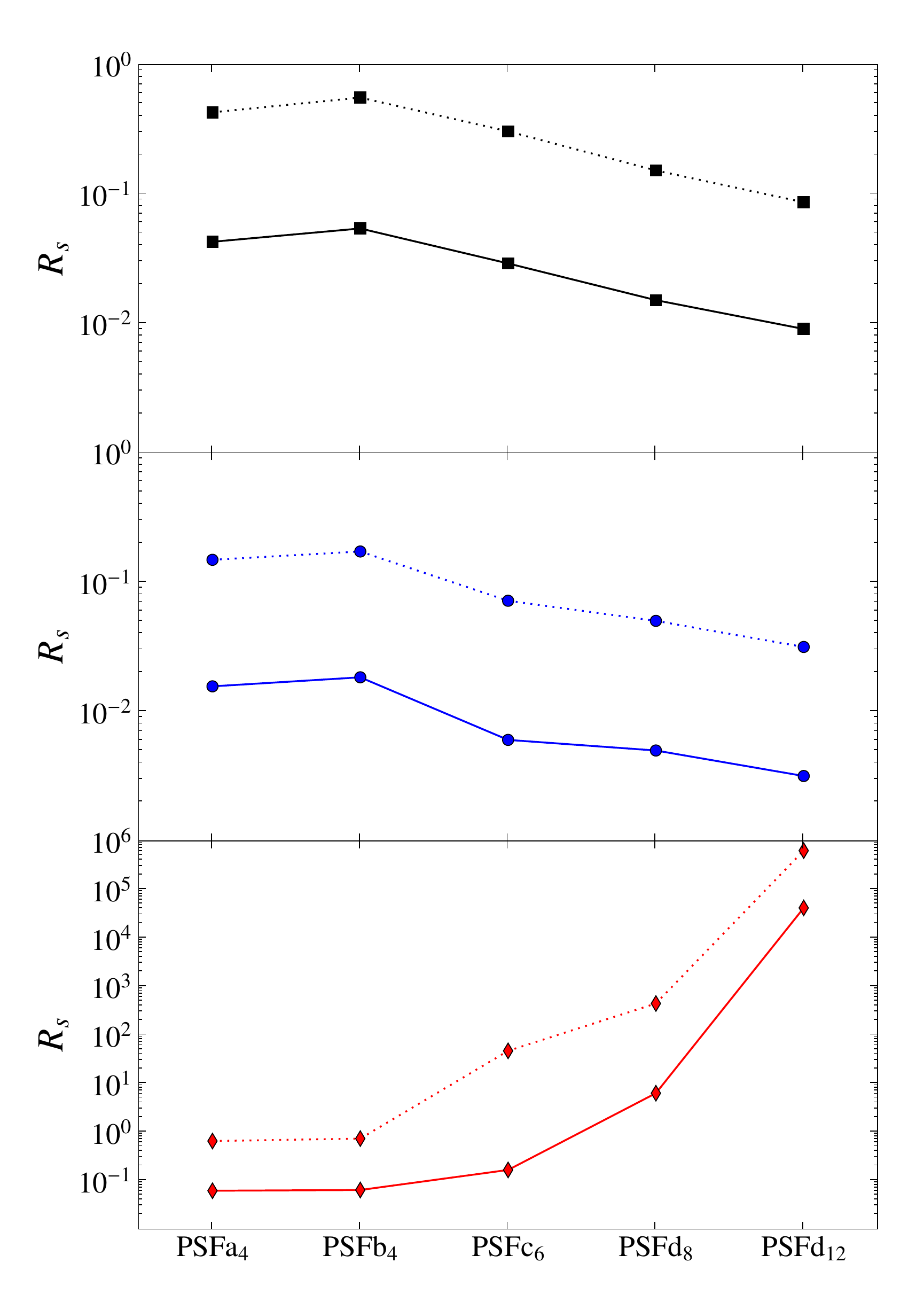}
 \caption{Distance in shapelet space $R_s$ between the mean deconvolved and the true intrinsic coefficients in dependence of the PSF model for the three methods {\sc Full} (top panel), {\sc Signific} (middle panel) and {\sc Same} (bottom panel). The mean is computed by averaging over all available values of $\alpha$ and $\beta$ (in total 60 independent combinations). Shown are the results for the moderate-noise simulations (solid line) and the high-noise simulations (dotted line).}
 \label{fig:high-noise_R}
\end{figure}
\begin{figure}[t]
 \includegraphics[width=\linewidth]{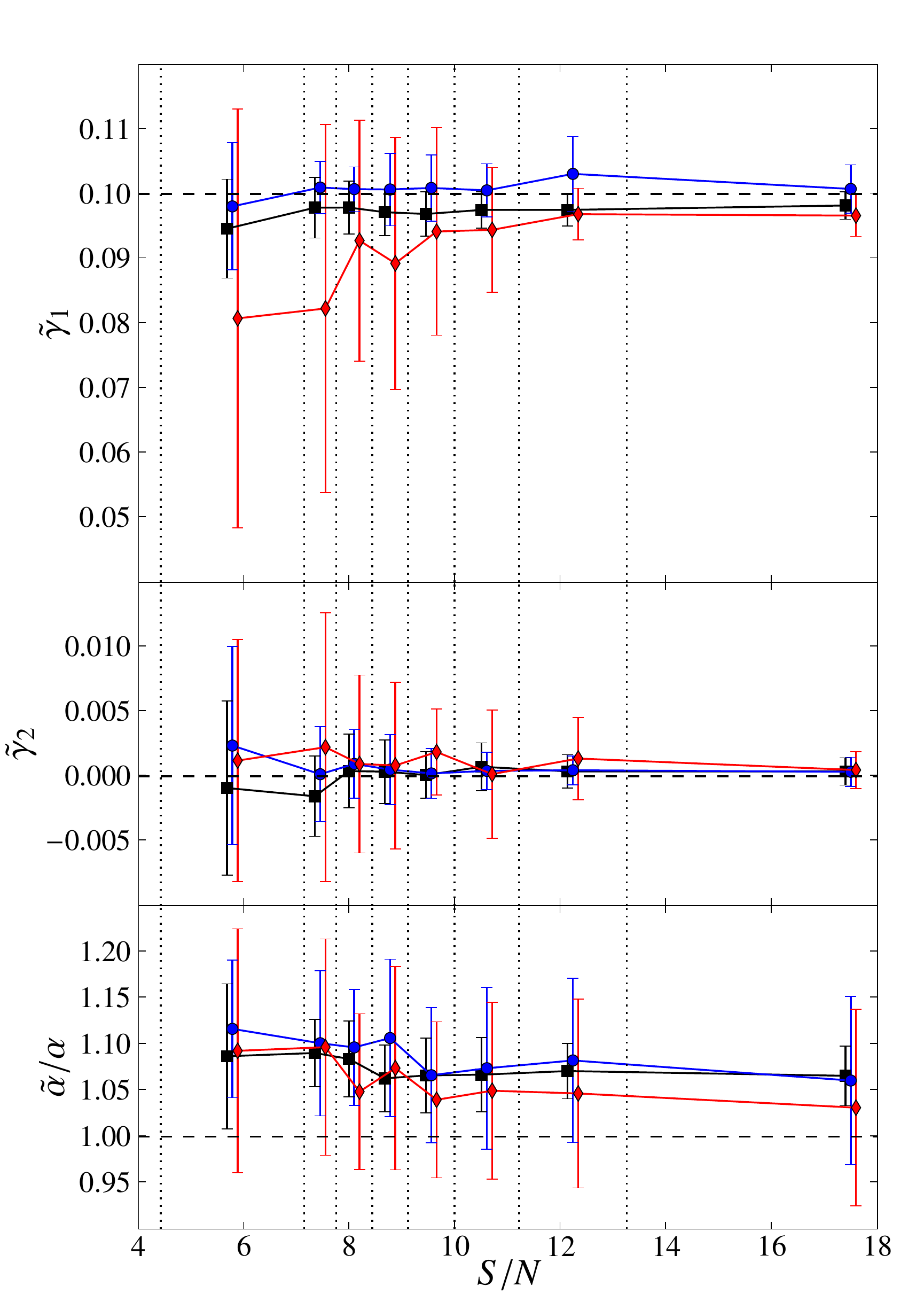}
 \caption{Recovered shear $\tilde{\vec\gamma}$ and intrinsic scale size $\tilde\alpha$ (in units of the true scale size $\alpha$) in the high-noise simulations in dependence of the $S/N$ of the convolved galaxy. The binning (dotted lines) is defined by the octiles of the $S/N$ distribution, therefore all bins contain the mean values of approx. 7 combinations of $\alpha$, $\beta$ for each PSF model, in total $\approx35$ independent settings. The data are plotted at the center of the bins and the methods are slightly offset horizontally for visualization purposes. Color code is as explained in Fig. \ref{fig:moderate-noise_average}.}
 \label{fig:high-noise_SN}
\end{figure}

Fig. \ref{fig:high-noise_R} confirms that as long as the kernel order is small, all three method perform quite similarly. But when the kernel order increases, {\sc Same} tries to recover a quadratically increasing number of deconvolved coefficients whose individual significance is lowered at the same time. On the other hand, {\sc Full} and {\sc Signific} make use of the redundancy of the overdetermined coefficient set, which is created by applying a rectangular matrix $P$ in \veqref{eq:shapelet_convolution}. As a direct consequence of computing the least-squares solution via $P^\dagger$, the higher the number of convolved coefficients and the lower the number of significant intrinsic coefficients, the better these intrinsic coefficients can be recovered from noisy measurements. This explains the decrease of $R_s$ with the kernel order for these two methods.

However, {\sc Full} does not perform perfectly as well. The bottom panel of Fig. \ref{fig:high-noise_average} reveals a bias on $\tilde{\alpha}$, independent of the PSF model. The reason for this is again overfitting. As {\sc Full} goes to higher orders than {\sc Signific} and {\sc Same}, it is even more affected by the pixel noise. As the decomposition determines $\gamma$ by minimizing $\chi^2$, $\gamma$ tends to become larger because this allows the model to fit a larger (increasingly noise-dominated) area, which reduces the overall residuals and thus $\chi^2$. {\sc Signific} and {\sc Same} behave similarly when the kernel order -- and hence the decomposition order -- becomes larger.

In order to prevent the shapelet models from creeping into the noisy areas around the object, it seems useful to constrain $\gamma$ not only from below but also from above. In addition to a guess on $\tilde{n}_{max}^f$, we therefore impose a constraint $\beta<\gamma<\sqrt{\beta^2+\alpha_{max}^2}$. Inferring both should be feasible when investigating observational data.

As our simulations comprise galaxy models of varying $S/N$ -- the models for both $f$ and $g$ have unit flux, so the surface brightness of the convolved object $h$ depends on $\alpha$ and $\beta$ --, it is illustrative to present the deconvolution results in $S/N$ bins.
Fig. \ref{fig:high-noise_SN} confirms that the two methods we propose here are very robust against image degradation. This is remarkable as many weak-lensing pipelines (and also {\sc Same} in this paper) suffer from an underestimation of the shear, which becomes increasingly prominent with decreasing $S/N$ \citep{Massey07.1}. Our statement from above, in which we related this drop to the high number of insignificant coefficients obtained from a deconvolution using {\sc Same}, is further supported by this figure. It is obvious that -- independent of the kernel model -- a low $S/N$ in pixel space results in a low $S/N$ in shapelet space. By obtaining the least-squares solution for the $f_\mathbf{n}$, {\sc Full} and {\sc Signific} boost the significance of the recovered coefficients and thus perform better in the low $S/N$ regime. The reason why $\tilde{\gamma}_1$ from {\sc Full} is consistently but insignificantly lower than the estimates from {\sc Same} is still somewhat unclear. A possible reason is the generally higher number of shapelet coefficients $h_\mathbf{n}$ for {\sc Full} and thus a more noticeable noise contamination.

\section{Conclusions}
\label{sec:conclusions}
Based on an analytic consideration of shapelet convolution, we have studied algorithms for the PSF deconvolution of galaxy images in shapelet space. The starting point are sum rules for shapelet convolution, showing that the intrinsic shapelet orders of PSF and image add in the convolved image, and that the squares of their scales are also added. We suggest an algorithm respecting these sum rules as well as possible in presence of noise, whose central step is the deconvolution of the convolved image with the pseudo-inverse of the convolution matrix. Applications to simulated images have shown that our algorithm performs very well and in many cases noticeably better than previously suggested methods. 
We identify three main reasons for the improved performance:
\begin{itemize}
\item As the sum rule for the shapelet order shows, the convolution transports power to higher shapelet modes. The mean signal-to-noise ratio of the convolved coefficients is thus reduced. Our reduction of the order during the deconvolution increases the signal-to-noise without the need of calibration.
\item The sum rules typically require a much higher shapelet order than the $\chi^2$ minimization, in particular if the PSF model is structured. Many of the high-order coefficients are thus highly insignificant. The significance of the coefficients is re-established by the reduction to the order $\bar{n}_{max}^f$, which is chosen such that the shapelet expansion contains only significant coefficients. Fortunately, $\bar{n}_{max}^f$ depends mainly on the signal-to-noise ratio of the galaxy, but only weakly: many different galaxy shapes can be deconvolved with the same $\bar{n}_{max}^f$. Thus, binning the galaxies into broad signal-to-noise bins will suffice.
\item The STEP-2 project \citep{Massey07.1} has shown that shear measurements generally depend strongly on the galaxy brightness. Our algorithm seems to have the advantage of lowering the influence of pixel noise in a well-defined manner.
\end{itemize}
We shall proceed to study the performance of our algorithm in shear measurements under realistic conditions.

\section*{Acknowledgments}
PM wants to thank Thomas Erben, Alex Boehnert, Ludovic van Waerbeke and Marco Lombardi for very fruitful discussions. PM is supported by the DFG Priority Programme 1177. MM is supported by the Transregio-Sonderforschungsbereich TRR 33 of the DFG.

\appendix
\section{Convolution scale and order}
\label{sec:proof}
For simplicity we restrict ourselves to the one-dimensional case. From \veqsref{eq:decomposition}{eq:convolution} it is apparent that 
\begin{equation}
\label{eq:conv2}
h(x)=\sum_{m,l}f_m g_l \int dx^\prime B_m(x^\prime;\alpha) B_l(x-x^\prime;\beta).
\end{equation}
We define $I_{m,l}(x;\alpha,\beta)$ as the integral in \veqref{eq:conv2} and decompose it into shapelets with scale size $\gamma$ and maximum order $N$,
\begin{equation}
I_{m,l}(x;\alpha,\beta) = \sum_n^N c_n B_n(x;\gamma).
\end{equation}
Considering \Veqsref{eq:shapelet_basis}{eq:dimless_basis}{eq:conv2}, we recognize that $N$ cannot be infinite but is determined by the highest modes of the expansions of $f$ and $g$, which we will call $M$ and $L$, respectively. Restricting to these modes and dropping all unnecessary constants, we can proceed, 
\begin{equation}
\begin{split}
&I_{M,L}(x;\alpha,\beta) =\\
&\int dx^\prime (x^\prime )^M \exp\left[-\frac{x^{\prime 2}}{2\alpha^2}\right] (x-x^\prime )^L \exp\left[-\frac{(x-x^\prime )^2}{2\beta^2}\right]=\\
&\sum_{i=0}^L (-1)^{L+1}\left(\begin{array}{c} L \\ i \end{array}\right)x^{L-i}\int dx^\prime (x^\prime)^{M+i} \exp\left[-\frac{(x-x^\prime)^2}{2\beta^2}-\frac{(x^\prime)^2}{2\alpha^2}\right],
\end{split}
\end{equation}
where we expanded $(x-x^\prime)^L$ in the last step. By employing \veqref{eq:natural_choice} and substituting $\tilde{x}=x^\prime-\frac{\alpha^2}{\gamma^2}x$, we can split the exponential,
\begin{equation}
\begin{split}
I_{M,L}(x;\alpha,\beta) =\sum_{i=0}^L& (-1)^{L+1}\left(\begin{array}{c} L \\ i \end{array}\right)x^{L-i}\exp\left[-\frac{x^2}{2\gamma^2}\right]\times\\
&\int d\tilde{x}\,\left(\tilde{x}+\frac{\alpha^2}{\gamma^2}x\right)^{M+i} \exp\left[-\frac{\gamma^2}{2\alpha^2\beta^2}\tilde{x}^2\right].
\end{split}
\end{equation}
Again, we expand $\left(\tilde{x}+\frac{\alpha^2}{\gamma^2}x\right)^{M+i}$, which yields the desired expression
\begin{equation}
\label{eq:conv_final}
\begin{split}
I_{M,L}(x;\alpha,\beta) =\sum_{i=0}^L& (-1)^{L+1}\left(\begin{array}{c} L \\ i \end{array}\right)\sum_{j=0}^{M+i}\frac{\alpha^{2(M+i-j)}}{\gamma^{M-L+2i-j}}\left(\begin{array}{c} M+i \\ j \end{array}\right)\, C_j\,\times\\
&\left(\frac{x}{\gamma}\right)^{M+L-j}\exp\left[-\frac{x^2}{2\gamma^2}\right],
\end{split}
\end{equation}
where we inserted $C_j \equiv \int d\tilde{x}\,\tilde{x}^j \exp\left[-\frac{\gamma^2}{2\alpha^2\beta^2}\tilde{x}^2\right]$. Apart from the omitted constants, the second line of \ref{eq:conv_final} is the definition of $B_{M+L-j}(x;\gamma)$ (cf. \veqsref{eq:shapelet_basis}{eq:dimless_basis}) which shows that the natural choice is well motivated. Moreover, as $j$ runs from 0 to $M+i$, we see that the maximum order $N$ is indeed $M+L$, as we have claimed in \veqref{eq:conv_order}. Since $C_j=0$ if $j$ is odd, the only states with non-vanishing power have the same parity as $M+L$.

In summary, the natural choice is inherited from the Gaussian weighting function in \veqref{eq:dimless_basis} and the maximum order the result of a product of polynomials.

\bibliography{../references}

\end{document}